\documentclass[11pt,a4paper]{article}
\usepackage{amsmath}
\usepackage[mathcal]{eucal}
\usepackage{latexsym}
\usepackage{amssymb}
\begin{titlepage} 
\title{Four dimensional gravity as an almost Poisson system}
\author{Eyo Eyo Ita III}
\medskip  
\input amssym.def
\input amssym.tex

\def \in{\indent}

\begin{document}  
\maketitle
\bigskip
\centerline{Physics Department, US Naval Academy} 
\smallskip
\centerline{Annapolis, Maryland}
\smallskip
\centerline{ita@usna.edu}
           
\bigskip    
                  
\begin{abstract}
In this paper we examine the phase space structure of a noncanonical formulation of 4-dimensional gravity referred to as the Instanton representation of Plebanski gravity (IRPG).  The typical Hamiltonian (symplectic) approach leads to an obstruction to the definition of a symplectic structure on the full phase space of the IRPG.  We circumvent this obstruction, using the Lagrange equations of motion, to find the appropriate generalization of the Poisson bracket.  It is shown that the IRPG does not support a Poisson bracket except on the vector constraint surface.  Yet there exists a fundamental bilinear operation on its phase space which produces the correct equations of motion and induces the correct transformation properties of the basic fields.  This bilinear operation is known as the almost-Poisson bracket, which fails to satisfy the Jacobi identity and in this case also the condition of antisymmetry.  We place these results into the overall context of nonsymplectic systems. 
\end{abstract}
\end{titlepage}

\section{Introduction}

The physics literature is replete with examples of dynamical systems whose Hamiltonian formulations admit a symplectic structure \cite{REPLETE}, \cite{REPLETE1}.  In certain systems a symplectic structure is not available, and one may instead have to work with a residual Poisson structure if one exists.  Such situations might occur when the phase space of the system is odd-dimensional, for example as in spin systems.  For systems lacking a Poisson structure, another possibility is that they might exhibit an almost Poisson structure.  An almost Poisson bracket satisfies all the usual properties of a Poisson bracket with the exception of the Jacobi identity, which is an integrability condition.  Almost Poisson structures in classical mechanics have had a long history, and typically have been associated with systems having nonholonomic constraints \cite{NONHOLONOMIC}, \cite{NONO}, \cite{NONO1}.  Some examples of almost-Poisson systems include autoparallels on Riemann--Cartan spacetime \cite{NONHOLONOMIC1}, the Foucault pendulum and the rolling sphere, and a point particle coupled to a magnetic monopole \cite{FEDOSOV}.\par 
\indent 
In \cite{EYO} a reformulation of gravity was introduced, referred to as the instanton representation of Plebanski gravity.  The instanton representation, with starting action $I_{Inst}$, uses a different set of variables than the conventional canonical descriptions of gravity.  The purpose of the present paper is to classify the Poisson structure which governs the evolution equations of $I_{Inst}$, and to determine the fundamental bilinear operation on its phase space variables.  A natural question is why one should want to study the structure of yet another formalism of gravity, when for instance one has the Ashtekar variables \cite{ASH2},\cite{ASH3} which provide a simple canonical structure.  It has been shown in \cite{EYO} that the evolution and constraint equations of $I_{Inst}$ contain comparatively few spatial derivatives acting on the Lagrange multiplier fields, which suggests that they could be more amenable to gauge-fixing procedures.  Given this prospect, it makes sense to learn as much as possible about the internal structure of $I_{Inst}$.  In this paper we would like to determine what sort of symplectic structure $I_{Inst}$ possesses.  But a first order of business will be to show that $I_{Inst}$ does indeed describe gravity for spacetimes of Pertov Types I, D and O.\par 
\indent 
The results of this paper will show that the instanton representation can best be thought of as an almost-Poisson system.  It is well-known that given a physical system with a complicated noncanonical structure, it is always possible to introduce auxiliary fields as necessary put the system into canonical form.  But that is not the purpose of this paper.  Rather, we would like to determine the correct structure of $I_{Inst}$ in terms of its original variables, without the introduction of auxiliary fields.  The motivation comes from the need to verify that $I_{Inst}$ is an internally consistent, stand-alone theory of gravity.\par 
\indent
The organization of this paper will be as follows.  In section 2 we provide a basic review of Poisson brackets for symplectic manifolds.  In section 3 we introduce the action for $I_{Inst}$ and attempt to identify its Poisson structure.  We find that the symplectic matrix, while nondegenerate, is too difficult to invert owing to the presence of spatial derivative operators which cannot be decoupled from the individual one forms in the wedge product.  In order to obtain a well-defined symplectic two form where the derivatives are decoupled, we must impose a constraint at the level of the starting action.  But this makes the symplectic two form degenerate, leading to an impasse.\par 
\indent  
In sections 4 and 5 we find the bilinear brackets directly by using the Lagrange equations for $I_{Inst}$, regarded as more fundamental.  In section 7 we evaluate the brackets against the conditions required for a Poisson structure.  The interpretation is that the action $I_{Inst}$ degenerates from a genuine symplectic structure into a Poisson structure which is not symplectic when the aformentioned constraint holds.  The structure derived directly from the equations of motion, however, is not a Poisson structure since it fails to satisfy the condition of antisymmetry and the Jacobi identity.  On the aforementioned constraint surface, this latter structure also becomes a Poisson structure which is not symplectic.  Section 7 is a conclusion section, including directions for future research.\par 
\indent 
Finally, prior to proceeding, possible extension of the formalism of this paper to third quantization as studied in \cite{THIRD},\cite{THIRD1},\cite{THIRD2},\cite{THIRD3}, \cite{THIRD4}, could constitute a basis for future potential research.

\section{Background and basic terminology}

Let $M$ be a manifold and let $f(M)$ denote the set of smooth functions on $M$.  A bilinear bracket operation $\{,\}$ from $f(M)\times{f}(M)\rightarrow{f}(M)$ satisfying the conditions of antisymmetry, bilinearity, the Leibniz rule and the Jacobi identity
\begin{align}
&(i)~\hbox{Antisymmetry}:~\{f,g\}=-\{g,f\};\\
&(ii)~\hbox{Bilinearity}:~\{\alpha{f}+\beta{g},h\}=\alpha\{f,h\}+\beta\{g,h\},~\alpha,\beta\in{C};\\
&(iii)~\hbox{Leibniz~rule}:~\{f,gh\}=\{f,g\}h+g\{f,h\};\\
&(iv)~\hbox{Jacobi~identity}:~\{f,\{g,h\}\}+\{g,\{h,f\}\}+\{h,\{f,g\}\}=0
\end{align}
\noindent
is known as a Poisson bracket, and $M$ is known as a Poisson manifold.  Let $\xi^i$, where $i=1,2,\dots{dim(M)}$ be local coordinates on $M$.  Then from the Poisson brackets one may define a Poisson tensor $P$ with components
\begin{eqnarray}
\label{TENSOR}
\{\xi^i,\xi^j\}=P^{ij}.
\end{eqnarray}
\noindent
The fundamental Poisson bracket (\ref{TENSOR}) induces the following bracket between any two smooth functions $f,g\in{C}^{\infty}(M)$
\begin{eqnarray}
\label{TENSOR1}
\{f,g\}=P^{ij}{{\partial{f}} \over {\partial\xi^i}}{{\partial{g}} \over {\partial\xi^j}}.
\end{eqnarray}
\noindent
As a consequence of the Jacobi identity, the Poisson tensor must satisfy the following relation
\begin{eqnarray}
\label{TENSOR2}
P^{lk}{{\partial{P}^{ij}} \over {\partial\xi^l}}+P^{li}{{\partial{P}^{jk}} \over {\partial\xi^l}}+
P^{lj}{{\partial{P}^{ki}} \over {\partial\xi^l}}=0.
\end{eqnarray}
\noindent
If $dim(M)=2n$, e.g. $M$ is of even dimension and if $P$ is nondegenerate in the sense that $Rank(P)=2n$, then there exists a fundamental two form $\Omega$ on $M$ such that $\Omega=P^{-1}$.  In local coordinates this is given by
\begin{eqnarray}
\label{TENSOR3}
\Omega={1 \over 2}\Omega_{ij}{d\xi^i}\wedge{d\xi^j}.
\end{eqnarray}
\noindent
If $d\Omega=0$, then this combined with the fact that $\Omega$ is nondegenerate endows $M$ with the structure of a symplectic manifold, and (\ref{TENSOR3}) is known as a symplectic two form.  Note that the condition that $\Omega$ be closed is equivalent to condition (\ref{TENSOR2}), which is due to the Jacobi identity.\par 
\indent
When one has a genuine symplectic manifold $M$, then one can define canonical transformations as the set of diffeomorphisms of $M$ which preserves the symplectic two form $\Omega$.  Such transformations are generated by vector fields $X$ such that $L_X\Omega=0$, where $L_X$ is the Lie derivative of $\Omega$ in the direction of $X$.  By Cartan's magic formula we have
\begin{eqnarray}
\label{CARTANMAGIC}
L_X\omega=d(i_X\Omega)+i_X(d\Omega)=0,
\end{eqnarray}
\noindent
where $i_X$ denotes the interior product in the direction of $X$.  When $M$ is symplectic, then the symplectic 2 form $\Omega$ is closed, and $d\Omega=0$.  From (\ref{CARTANMAGIC}) this implies that $d(i_X\Omega)=0$, and for manifolds with trivial cohomology, the Poincare lemma implies that
\begin{eqnarray}
\label{CARTANMAGIC1}
i_X\omega=df
\end{eqnarray}
\noindent
for some function $f$ on $M$.  Hence, for symplectic manifolds there is a correspondence between functions $f$ and vector fields $X$.\par 
\indent
However, if the bilinear operation $\{,\}$ on $M$ satisfies conditions (i), (ii) and (iii) but does not satisy condition (iv), the Jacobi identity, then $\{,\}$ is known as an almost-Poisson bracket and $M$ is an almost-Poisson manifold.  For almost-Poisson manifolds, given that there is no closed symplectic 2 form, the question arises as to whether there exists an analogous correspondence between phase space functions and vector fields in some suitable sense.  We will relegate this question to \cite{POISSON}.\footnote{For clarification, the brackets which we will derive for gravity will satisfy neither the condition of antisymmetry nor the Jacobi identity and do not meet the criteria of an almost-Poisson structure in the strict sense.  We use the term almost-Poisson more-so in the context that the system starts as a non-Poisson system and in the end, when a certain constraint holds, becomes Poisson system.}

\section{Ashtekar variables and the instanton representation evolution equations}

Let $M$ be a four dimensional spacetime manifold of topology $M=\Sigma\times{R}$, for some 3 dimensional spatial manifold $\Sigma$ of a given topology.  Then the action for General Relativity in the Ashtekar variables is given by \cite{ASH2}, \cite{ASH3}
\begin{eqnarray}
\label{STARTING}
I=\int{dt}\int_{\Sigma}d^3x\biggl[\widetilde{\sigma}^i_a\dot{A}^a_i+A^a_0D_i\widetilde{\sigma}^i_a-\epsilon_{ijk}\widetilde{\sigma}^i_aB^j_aN^k
-iN\sqrt{\hbox{det}\widetilde{\sigma}}\bigl(\Lambda+B^i_a(\widetilde{\sigma}^{-1})^a_i\bigr)\biggr],
\end{eqnarray}
where the phase space variables are a self-dual $SO(3)$ gauge connection $A^a_i$,\footnote{For index conventions, symbols from the beginning of the Latin alphabet $a,b,c,\dots$ refer to internal $SO(3)$ indices, while those from the middle $i,j,k,\dots$ refer to spatial indices} (with magnetic field $B^i_a$) and a densitized triad $\widetilde{\sigma}^i_a$.  These phase space variables $(A^a_i,\widetilde{\sigma}^i_a)$ form a canonical pair, and the fields $A^a_0,N^i,N$ are auxiliary fields.\par 
\indent 
The Lagrange equations of (\ref{STARTING}) consist of the constraints,
\begin{eqnarray} 
\label{COONSTRAINT}
\frac{\delta{I}}{\delta{A}^a_0}=D_i\widetilde{\sigma}^i_a=0;~~\frac{\delta{I}}{\delta{N}^i}=\epsilon_{ijk}\widetilde{\sigma}^i_aB^j_aN^k;\nonumber\\
\frac{\delta{I}}{\delta{N}}=\sqrt{\hbox{det}\widetilde{\sigma}}\bigl(\Lambda+B^i_a(\widetilde{\sigma}^{-1})^a_i\bigr)=0,  
\end{eqnarray}
\noindent 
and the evolution equations for the dynamical variables
\begin{eqnarray}
\label{DERIVATION}
\dot{A}^a_i=D_iA^a_0+\epsilon_{ijk}B^j_aN^k-iN\sqrt{\hbox{det}\widetilde{\sigma}}B^n_d(\widetilde{\sigma}^{-1})^d_i(\widetilde{\sigma}^{-1})^a_n,
\end{eqnarray} 
\noindent 
and 
\begin{eqnarray} 
\label{DERIVATION1}
\dot{\widetilde{\sigma}^i_a}=f_{abc}\widetilde{\sigma}^i_bA^c_0+\epsilon_{mjk}\epsilon^{jni}D_n(\widetilde{\sigma}^m_aN^k)
-i\epsilon^{ijk}D_j(N\sqrt{\hbox{det}\widetilde{\sigma}}(\widetilde{\sigma}^{-1})^a_k).
\end{eqnarray}
\noindent 
The canonical analysis of (\ref{STARTING}) has been treated in the literature.  In this paper we will introduce a noncanonical formulation of gravity, which exhibits a new kind of Poisson structure.  Let us perform the following transformation 
\begin{eqnarray} 
\label{DERIVATION2}
\widetilde{\sigma}^i_a=\Psi_{ae}B^i_e,~\hbox{det}B,~\hbox{det}\Psi\neq{0}
\end{eqnarray}
\noindent 
to a new set of variables.  In this paper we will write down an action $I[A,\Psi]$ corresponding to the variables $(A^a_i,\Psi_{ae})$ and determine its Poisson structure.\par 
\indent   
It may be tempting to perform the substitution (\ref{DERIVATION2}) at the level of the action (\ref{STARTING}).  We will show that this procedure does in fact lead to the correct action by showing that the substitution (\ref{DERIVATION2}) commutes with the equations of motion.  That is, implementing (\ref{DERIVATION2}) at the level of the equations of motion derived from (\ref{STARTING}), which is defined on the phase space $(A^a_i,\widetilde{\sigma}^i_a)$, will yield the same equations as the equations of motion resulting from (\ref{DERIVATION2}) made at the level of the action (\ref{STARTING}), transforming it into an action $I[A,\Psi]$.\par 
\indent 
Substitution of (\ref{DERIVATION2}) into (\ref{DERIVATION}) yields the following equation for $A^a_i$ on the phase space $A,\Psi$
\begin{eqnarray} 
\label{DERIVATION3}
\dot{A}^a_i=D_iA^a_0+\epsilon_{ijk}B^j_aN^k-iN\sqrt{\hbox{det}B}\sqrt{\hbox{det}\Psi}(\Psi^{-1}\Psi^{-1})^{ad}(B^{-1})^d_i  
\end{eqnarray} 
\noindent 
Now we will derive the evolution equation for $\Psi$.  Applying the Leibniz rule, the time derivative of (\ref{DERIVATION2}) is given by
\begin{eqnarray} 
\label{DERIVATION4}
\dot{\widetilde{\sigma}^i_a}=\dot{\Psi}_{ae}B^i_e+\Psi_{ae}\dot{B}^i_e=\dot{\Psi}_{ae}B^i_e+\Psi_{ae}\epsilon^{ijk}D_j\dot{A}^e_k.
\end{eqnarray}
\noindent 
Substituting (\ref{DERIVATION}) into (\ref{DERIVATION4}) we have\footnote{We will substitute (\ref{DERIVATION2}) near the end of the derivation, for convenience.}
\begin{eqnarray} 
\label{DERIVATION5}
\dot{\widetilde{\sigma}^i_a}=\dot{\Psi}_{ae}B^i_e+\Psi_{ae}\epsilon^{ijk}D_j\Bigl(D_kA^e_0+\epsilon_{kmn}B^m_eN^n-iN\sqrt{\hbox{det}\widetilde{\sigma}}
\Psi_{ae}B^n_d(\widetilde{\sigma}^{-1})^d_k(\widetilde{\sigma}^{-1})^e_n\Bigr)\nonumber\\
=\dot{\Psi}_{ae}B^i_e+\Psi_{ae}f^{ebg}B^i_bA^g_0
+\Psi_{ae}\epsilon^{ijk}D_j\Bigl(\epsilon_{kmn}B^m_eN^n-iN\sqrt{\hbox{det}\widetilde{\sigma}}
\Psi_{ae}B^n_d(\widetilde{\sigma}^{-1})^d_k(\widetilde{\sigma}^{-1})^e_n\Bigr)
\end{eqnarray}
\noindent 
where we have used $\epsilon^{ijk}D_jD_kA^e_0=f^{ebg}B^i_bA^g_0$, namely the definition of curvature as the commutator of two covariant derivatives.\par 
\indent 
Next we will apply the Leibniz rule to the last terms of (\ref{DERIVATION5}), bringing $\Psi_{ae}$ into the large brackets and subtracting the remainder.  This yields
\begin{eqnarray} 
\label{DERIVATION6}
\dot{\widetilde{\sigma}^i_a}=\dot{\Psi}_{ae}B^i_e+\Psi_{ae}f^{ebg}B^i_bA^g_0\nonumber\\
+D_j\Bigl(\epsilon^{ijk}\epsilon_{kmn}\Psi_{ae}B^m_eN^n-i\epsilon^{ijk}N\sqrt{\hbox{det}\widetilde{\sigma}}\Psi_{ae}B^n_d(\widetilde{\sigma}^{-1})^d_k(\widetilde{\sigma}^{-1})^e_n\Bigr)\nonumber\\
-\Bigl(\epsilon^{ijk}\epsilon_{kmn}B^m_eN^n-i\epsilon^{ijk}N\sqrt{\hbox{det}\widetilde{\sigma}}B^n_d(\widetilde{\sigma}^{-1})^d_k(\widetilde{\sigma}^{-1})^e_n\Bigr)D_j\Psi_{ae}.
\end{eqnarray}
\noindent 
So we have two expressions for $\dot{\widetilde{\sigma}^i_a}$, namely (\ref{DERIVATION1}) and (\ref{DERIVATION6}), which we can set equal to each other.  Using (\ref{DERIVATION2}) in the 
middle line of (\ref{DERIVATION6}) and in the first term of (\ref{DERIVATION1}) and upon relabelling of indices, one sees that the second and third terms on the right hand side of (\ref{DERIVATION1}) are the same as the middle line of (\ref{DERIVATION6}).  So cancelling these terms out, we are left with the following relation
\begin{eqnarray} 
\label{DERIVATION7}
f_{abc}\Psi_{be}B^i_eA^c_0=\dot{\Psi}_{ae}B^i_e+\Psi_{ae}f^{ebc}B^i_bA^c_0\nonumber\\
-\bigl(\delta^i_m\delta^j_n-\delta^i_n\delta^j_m\bigr)B^m_eN^nD_j\Psi_{ae}+iN\sqrt{\hbox{det}\widetilde{\sigma}}\epsilon^{ijk}(\Psi^{-1})^{ed}(\widetilde{\sigma}^{-1})^d_kD_j\Psi_{ae}.
\end{eqnarray}
\noindent 
Multiplying (\ref{DERIVATION7}) by $(B^{-1})^g_i$ and rearranging, we get the following evolution equation for $\Psi_{ag}$
\begin{eqnarray} 
\label{DERIVATION8}
\dot{\Psi}_{ag}=\bigl(f_{abc}\Psi_{bg}+f_{gbc}\Psi_{ab}\bigr)A^c_0+N^jD_j\Psi_{ag}-N^i(B^{-1})^g_iB^j_eD_j\Psi_{ae}\nonumber\\
+iN\sqrt{\frac{\hbox{det}\Psi}{\hbox{det}B}}\epsilon^{fbg}(\Psi^{-1}\Psi^{-1})^{fe}B^j_bD_j\Psi_{ae}.
\end{eqnarray}
\noindent 
Having derived the evolution equations for $A^a_i$ and $\Psi_{ae}$, we will now propose a starting action for GR based on these variables, as a point of departure for this paper.

\section{Poisson structure of $I_{Inst}$}

We propose the following first order action, referred to as the instanton representation of Plebanski gravity \cite{EYO}, as a starting action for GR based on the noncanonical variables $(A^a_i,\Psi_{ae})$ 
\begin{eqnarray}
\label{CANON}
I_{Inst}=\int{dt}\int_{\Sigma}d^3x\biggl[\Psi_{ae}B^i_e\dot{A}^a_i+A^a_0B^i_eD_i\Psi_{ae}\nonumber\\
-\epsilon_{ijk}N^iB^j_eB^k_a\Psi_{ae}-N(\hbox{det}B)^{1/2}\sqrt{\hbox{det}\Psi}\bigl(\Lambda+\hbox{tr}\Psi^{-1}\bigr)\biggr],
\end{eqnarray}
\noindent
where $\Lambda$ is the cosmological constant.  The basic phase space variables are a $SO(3)$ connection $A^a_i$ and a three by three matrix $\Psi_{ae}$, taking its values in two copies of $SO(3)$, and $N$, $N^i$ and $A^a_0$ are auxiliary fields.\par 
\indent   
Equation (\ref{CANON}) is the result of substitution of (\ref{DERIVATION2}) into (\ref{STARTING}) which is not, strictly speaking in general, a valid procedure.\footnote{This is because the Lagrange's equations assume independent degrees of freedom.  When a substitution is made at the level of the action, it is possible that degrees of freedom may be constrained, resulting in a different theory than the original action.}  However, in this case the procedure works, which we will demonstrate by showing obtaining the same equations (\ref{COONSTRAINT}), (\ref{DERIVATION}) and (\ref{DERIVATION6}) from the action (\ref{CANON}).  Ultimately in this paper, we would like to determine the intrinsic structure of (\ref{CANON}).  The equations of motion for the auxiliary fields, for $(\hbox{det}B)\neq{0}$ and $(\hbox{det}\Psi)\neq{0}$, imply 
\begin{eqnarray} 
\label{AUXILIARY}
B^i_eD_i\Psi_{ae}=0;~~\epsilon_{dae}\Psi_{ae}=0;~~\Lambda+\hbox{tr}\Psi^{-1}=0,
\end{eqnarray}
\noindent 
which are the same result of substituting (\ref{DERIVATION2}) into the equations (\ref{COONSTRAINT}).  In the next section we will show that this feature extends to the evolution equations.\par 
\indent
In this paper we will attempt to unravel the Poisson and symplectic structures of (\ref{CANON}).  The symplectic potential from (\ref{CANON}) is given by 
\begin{eqnarray}
\label{ONEFORM}
\theta_{Inst}=\int_{\Sigma}d^3x\Psi_{ae}B^i_e\delta{A}^a_i.
\end{eqnarray}
\noindent
To find the symplectic 2-form $\Omega_{Inst}$, we will proceed by first taking the exterior functional derivative of (\ref{ONEFORM}), yielding
\begin{eqnarray}
\label{ONEFORM1}
\delta\theta_{Inst}=\int_{\Sigma}d^3xB^i_e{\delta\Psi_{ae}}\wedge{\delta{A}^a_i}
+\int_{\Sigma}d^3x\Psi_{ae}\epsilon^{ijk}{D_j(\delta{A}^e_k)}\wedge{\delta{A}^a_i}.
\end{eqnarray}
\noindent
The first term of (\ref{ONEFORM1}) is relatively straightforward to interpret.  However the second term will yield a symplectic matrix with derivatives acting on $\delta{A}^e_k$.  This could be problematic for finding the Poisson matrix, since one would need to invert a symplectic matrix containing differential operators, whose action moreover appears to be unclear.  Let us average this term with its clone with indices relabelled $a\leftrightarrow{e}$ and $i\leftrightarrow{k}$.  Hence the second term on the right hand side of (\ref{ONEFORM1}) is the same as
\begin{eqnarray}
\label{ONEFORM2}
{1 \over 2}\int_{\Sigma}d^3x\Psi_{ae}\epsilon^{ijk}{D_j(\delta{A}^e_k)}\wedge{\delta{A}^a_i}
+{1 \over 2}\int_{\Sigma}d^3x\Psi_{ea}\epsilon^{ijk}{\delta{A}^e_k}\wedge{D_j(\delta{A}^a_i)},
\end{eqnarray}
\noindent
where we have used the anticommutativity of one forms to anticommute $D_j(\delta{A}^a_i)$ to the right in the second term.  Let us now decompose $\Psi_{ae}=\Psi_{(ae)}+\Psi_{[ae]}$ into its symmetric and its antisymmetric parts.  Then (\ref{ONEFORM2}) splits into four terms
\begin{eqnarray}
\label{ONEFORM3}
{1 \over 2}\int_{\Sigma}d^3x\Psi_{(ae)}\epsilon^{ijk}{D_j(\delta{A}^e_k)}\wedge{\delta{A}^a_i}
+{1 \over 2}\int_{\Sigma}d^3x\Psi_{(ea)}\epsilon^{ijk}{\delta{A}^e_k}\wedge{D_j(\delta{A}^a_i)}+Q,
\end{eqnarray}
\noindent
where $Q$, which is linear in the antisymmetric part of $\Psi_{ae}$, is given by
\begin{eqnarray}
\label{ONEFORM4}
Q={1 \over 2}\int_{\Sigma}d^3x\Psi_{[ae]}\epsilon^{ijk}{D_j(\delta{A}^e_k)}\wedge{\delta{A}^a_i}
+{1 \over 2}\int_{\Sigma}d^3x\Psi_{[ea]}\epsilon^{ijk}{\delta{A}^e_k}\wedge{D_j(\delta{A}^a_i)}\nonumber\\
={1 \over 2}\int_{\Sigma}d^3x\Psi_{[ae]}\epsilon^{ijk}\bigl({D_j(\delta{A}^e_k)}\wedge{\delta{A}^a_i}
-{\delta{A}^e_k}\wedge{D_j(\delta{A}^a_i)}\bigr).
\end{eqnarray}
\noindent
Using $\Psi_{(ae)}=\Psi_{(ea)}$, the first two terms of (\ref{ONEFORM3}) can be combined via the Leibniz rule, followed by an integration by parts to transfer the derivatives from ${\delta{A}^e_k}\wedge{\delta{A}^a_i}$ onto $\Psi_{(ae)}$.\footnote{In this paper we will always assume that all boundary terms can be discarded, either by suitable fall-off conditions at infinity for all fields or by restriction to spatial manifolds $\Sigma$ having no boundary.  While the latter case poses no obvious difficulty, note that the former case does not preserve the conditions $\hbox{det}\Psi,\hbox{det}B\neq{0}$, hence violating the equivalence to General Relativity at infinity.  This is not an issue for $A,\Psi$ within the bulk of spacetime.}  The result of this is a well-defined two form
\begin{eqnarray}
\label{ONEFORM5}
{1 \over 2}\int_{\Sigma}d^3x\epsilon^{ijk}(D_j\Psi_{(ae)}){\delta{A}^a_i}\wedge{\delta{A}^e_k},
\end{eqnarray}
\noindent
where we have used anticommutativity of one forms.  Using these results we can now write (\ref{ONEFORM1}) in the following form
\begin{eqnarray}
\label{ONEFORM6}
\delta\theta_{Inst}=\Omega_{Inst}+W,
\end{eqnarray}
\noindent
where
\begin{eqnarray}
\label{ONEFORM7}
\Omega_{Inst}=\int_{\Sigma}d^3x\biggl[B^i_e{\delta\Psi_{(ae)}}\wedge{\delta{A}^a_i}
+{1 \over 2}\epsilon^{ijk}(D_j\Psi_{(ae)}){\delta{A}^a_i}\wedge{\delta{A}^e_k}\biggr];\nonumber\\
W=\int_{\Sigma}d^3x\biggl[B^i_e{\delta\Psi_{[ae]}}\wedge{\delta{A}^a_i}
+{1 \over 2}\Psi_{[ae]}\epsilon^{ijk}\bigl(D_j(\delta{A}^e_k)\wedge{\delta{A}^a_i}
-{\delta{A}^e_k}\wedge{D_j(\delta{A}^a_i)}\bigr)\biggr].
\end{eqnarray}
\noindent
Note that $\Omega_{Inst}$ has the clear interpretation as a symplectic two form on the phase space $\Psi_{(ae)},A^a_i$, whereas it appears difficult or impossible to decouple the spatial derivatives from the one forms $\delta{A}$ in $W$.\par 
\indent
If one is able to invert the symplectic matrix, then one could deduce the Poisson brackets for $I_{Inst}$.  But the contribution due to $W$ is ill-defined owing to these derivatives.  By choosing $\Psi_{[ae]}=0$ at the level of the action (\ref{CANON}), then we have $W=0$ and are left with a symplectic matrix with components\footnote{The factor of ${1 \over 2}$ has been extracted as in the definition of two forms.}
\begin{displaymath}
\Omega_{IJ}(x,y)=
\left(\begin{array}{cc}
\epsilon^{ijk}D_j\Psi_{ae} & -\delta_{ab}B^i_g\\
\delta_{ce}B^k_f & 0\\
\end{array}\right)
\delta^{(3)}(x,y)\biggl\vert_{\psi_{[ae]}=0},
\end{displaymath}
\noindent
which is clearly well-defined.\par 
\indent  
The original phase space $\Gamma_{Inst}$ was of dimension $9+9=18$ which is even.  But upon implementation of $\Psi_{[ae]}=0$, which incidentally is the vector constraint, the middle equation of (\ref{AUXILIARY}), then $\Psi_{ae}$ has three fewer degrees of freedom compared to $A^a_i$ and the phase space dimension becomes $9+6=15$ which is odd.  So it would be more appropriate at this stage to classify the manifold $A,\Psi$ as a Poisson manifold which is not symplectic, since the matrix $\Omega_{IJ}$ is now degenerate.  One possibility is to attempt to reduce the system, e.g. by some sort of gauge-fixing mechanism reduce the connection $A^a_i$ by three degrees of freedom to get a phase space of dimension $6+6=12$.  However we will postpone that approach for future research once further developed.  Since it is really the bilinear brackets that we are after, then we will rather find them more directly by appealing to the Lagrange equations for $I_{Inst}$, which are more straightforward to find.  Then we can determine, in retrospect, the precise nature of the true symplectic matrix if one exists.

\section{Lagrange's equations of motion}

The system defined by the action (\ref{CANON}) is noncanonical since $\Psi_{ae}$ is not the momentum conjugate to $A^a_i$.  We would like to determine their Poisson brackets but the symplectic matrix $\Omega_{IJ}$ is degenerate for $\Psi_{[ae]}=0$.  Off the constraint surface, we will describe (\ref{CANON}) analogously to a Hamiltonian system since using the Lagrange equations, we will be able to deduce $\dot{\Psi}_{ae},\dot{A}^a_i$ directly.  In the spirit of \cite{MHD}, we can generalize the Poisson bracket as necessary to produce these equations.  Assuming that the resulting Poisson matrix $P^{IJ}$ is invertible, then we can invert it and find the symplectic matrix indirectly.\footnote{This is mainly for comparison purposes with $\Omega_{IJ}$ from the previous section.  Upon finding the brackets themselves, we will have achieved the main aim of this paper.}  First let us re-write the action by integrating (\ref{CANON}) by parts
\begin{eqnarray}
\label{CANON9}
I_{Inst}=\int{dt}\int_{\Sigma}d^3x\biggl[\Psi_{ae}B^i_e\bigl(F^a_{0i}-\epsilon_{ijk}B^j_aN^k\bigr)\nonumber\\
-N(\hbox{det}B)^{1/2}\sqrt{\hbox{det}\Psi}\bigl(\Lambda+\hbox{tr}\Psi^{-1}\bigr)\biggr].
\end{eqnarray}
\noindent
Then the Lagrange equation of motion for $\Psi_{ae}$ is given by
\begin{eqnarray}
\label{CANON10}
{{\delta{I}_{Inst}} \over {\delta\Psi_{ae}}}
=B^i_e\bigl(F^a_{0i}-\epsilon_{ijk}B^j_aN^k\bigr)
+N(\hbox{det}B)^{1/2}\sqrt{\hbox{det}\Psi}(\Psi^{-1}\Psi^{-1})^{ea}=0,
\end{eqnarray}
\noindent
which we will intepret as an evolution equation for $A^a_i$.  Left-multiplication by $B^{-1}$ yields
\begin{eqnarray}
\label{CANON11}
F^a_{0i}-\epsilon_{ijk}B^j_aN^k+N(\hbox{det}B)^{1/2}\sqrt{\hbox{det}\Psi}(B^{-1})^e_i(\Psi^{-1}\Psi^{-1})^{ea}=0.
\end{eqnarray}
\noindent
Next, we will find the Lagrange equation for $A^d_m$.  The contribution from the $B^i_eF^a_{0i}$ term, after the appropriate integrations by parts, is given by
\begin{eqnarray}
\label{APPROPRIATE}
{\delta \over {\delta{A}^d_m}}\Bigl(\int_{\Sigma}d^3x\Psi_{ae}B^i_eF^a_{0i}\Bigr)\nonumber\\
={\delta \over {\delta{A}^d_m}}\Bigr(\int_{\Sigma}d^3x\Psi_{ae}B^i_e\bigl(\dot{A}^a_i-\partial_iA^a_0-f^{abc}A^b_iA^c_0\bigr)\Bigr)\nonumber\\
=\epsilon^{mji}D_j(\Psi_{ad}F^a_{0i})-{\partial \over {\partial{t}}}(\Psi_{de}B^m_e)-f^{dca}A^c_0(\Psi_{ae}B^m_e).
\end{eqnarray}
\noindent
The contributions from the $\vec{V}[\vec{N}]$ and $H[N]$ terms will be directly proportional to the $\vec{V}$ and $H$ constraints and their spatial derivatives and therefore vanish on the constraint surface.\footnote{It must subsequently be shown that these constraints form a closed algebra, an exercise which we will relegate to the second paper in this series \cite{POISSON}.}  The desired equation of motion is then the vanishing of (\ref{APPROPRIATE}).  Note, using the relation $D_0v_a=\partial_0v_a+f_{abc}A^b_0v_c$, that we can combine the last two terms of (\ref{APPROPRIATE}) into a temporal covariant derivative, hence an equation of motion
\begin{eqnarray}
\label{APPROPRIATE1}
\epsilon^{mij}D_j(\Psi_{ad}F^a_{0i})+D_0(\Psi_{de}B^m_e)=0.
\end{eqnarray}
\noindent
Let us expand (\ref{APPROPRIATE1}) using the Leibniz rule
\begin{eqnarray}
\label{APPROPRIATE2}
\epsilon^{mij}(D_j\Psi_{ad})F^a_{0i}+\epsilon^{mij}\Psi_{ad}D_jF^a_{0i}+(D_0\Psi_{de})B^m_e+\Psi_{de}D_0B^m_e=0.
\end{eqnarray}
To put (\ref{APPROPRIATE2}) into a more transparent form it will be convenient to recall the Bianchi identity which splits into temporal and spatial components
\begin{eqnarray}
\label{APPROPRIATE3}
\epsilon^{\mu\nu\rho\sigma}D_{\nu}F^e_{\rho\sigma}=0\longrightarrow{D}_iB^i_e=0;~~D_0B^i_e=\epsilon^{ijk}D_jF^e_{0k}.
\end{eqnarray}
\noindent
Substituting the second Bianchi identity of (\ref{APPROPRIATE3}) into the last term of (\ref{APPROPRIATE2}) and 
relabelling $k\rightarrow{i}$ on this term (and relabelling $a\rightarrow{e}$ on the first term, we get
\begin{eqnarray}
\label{APPROPRIATE4}
(D_0\Psi_{ae})B^m_e+\epsilon^{mij}F^e_{0i}(D_j\Psi_{ed})+(\Psi_{de}-\Psi_{ed})\epsilon^{mjk}D_jF^e_{0k}=0.
\end{eqnarray}
\noindent
Then using the definition $B^i_e={1 \over 2}\epsilon^{ijk}F^e_{jk}$ and multiplying by $B^{-1}$, equation (\ref{APPROPRIATE4}) then becomes
\begin{eqnarray}
\label{APPROPRIATE5}
D_0\Psi_{ae}+\epsilon^{mij}(B^{-1})^e_mF^g_{0i}(D_j\Psi_{ad})=-2\Psi_{[de]}(B^{-1})^e_m\epsilon^{mjk}D_jF^e_{0k}.
\end{eqnarray}
\noindent
It is satisfactory to implement the constraints at the level after, and not before, applying the equations of motion.  Therefore we will set the right hand side of (\ref{APPROPRIATE5}) to zero, via the vector constraint.

\section{Fundamental brackets of the basic fields}

We will now determine the basic brackets for $\Psi_{ae}$ and $A^a_i$ using their Lagrange's equations of motion.  Prior to proceeding we will make the replacement $\Psi_{(ae)}\rightarrow\Psi_{ae}$ as a weak equality.\footnote{This will be for calculational convenience of what follows.  This step will be justified in the end when we show that $\Psi_{[ae]}=0$ is preserved under time evolution.}  So (\ref{CANON11}) and (\ref{APPROPRIATE5}), repeated here for completeness, will form the starting point 
\begin{eqnarray}
\label{CANON24}
F^a_{0i}-\epsilon_{ijk}B^j_aN^k+N(\hbox{det}B)^{1/2}\sqrt{\hbox{det}\Psi}(B^{-1})^e_i(\Psi^{-1}\Psi^{-1})^{ea}=0;\nonumber\\
D_0\Psi_{ae}=-\epsilon^{ijk}(B^{-1})^e_iF^g_{0j}D_k\Psi_{ag}.
\end{eqnarray}
\noindent
The first line of (\ref{CANON24}) is an evolution equation for $A^a_i$.  To obtain an evolution equation for $\Psi_{ae}$ we substitute $F^g_{0j}$ from the first equation into the second equation, which yields
\begin{eqnarray}
\label{CANON25}
D_0\Psi_{ae}=-\epsilon^{ijk}(B^{-1})^e_i\epsilon_{jmn}B^m_gN^nD_k\Psi_{ag}\nonumber\\
+N(\hbox{det}B)^{1/2}\sqrt{\hbox{det}\Psi}\epsilon^{ijk}(B^{-1})^e_i(B^{-1})^f_j(\Psi^{-1}\Psi^{-1})^{fg}D_k\Psi_{ag}.
\end{eqnarray}
\noindent
Using epsilon identities and the definition of determinants, we have
\begin{eqnarray}
\label{CANON26}
D_0\Psi_{ae}=-\bigl(\delta^k_m\delta^i_n-\delta^k_n\delta^i_m\bigr)(B^{-1})^e_iB^m_gN^nD_k\Psi_{ag}\nonumber\\
+N(\hbox{det}B)^{-1/2}\epsilon^{efd}(\Psi^{-1}\Psi^{-1})^{fg}B^k_dD_k\Psi_{ag}\nonumber\\
=-N^i(B^{-1})^e_iB^k_gD_k\Psi_{ag}+N^kD_k\Psi_{ae}
+N(\hbox{det}B)^{-1/2}\epsilon^{efd}(\Psi^{-1}\Psi^{-1})^{fg}B^k_dD_k\Psi_{ag}.
\end{eqnarray}
Note that the first term on the right hand side of (\ref{CANON26}) is directly proportional to the Gauss' constraint, which we will set weakly to zero.  Using the relation
\begin{eqnarray}
\label{THERELATION}
D_0\Psi_{ae}=\dot{\Psi}_{ae}+A^b_0\bigl(f_{abc}\Psi_{ce}+f_{ebc}\Psi_{ac}\bigr), 
\end{eqnarray}
\noindent
then we can separate the part of (\ref{CANON26}) containing $\dot{\Psi}_{ae}$ from the remaining part, which involves a gauge transformation with parameter $A^b_0$.  The result is that the Lagrange equations for (\ref{CANON}) imply the evolution equations
\begin{eqnarray}
\label{CANON28}
\dot{A}^a_i=D_iA^a_0+\epsilon_{ijk}B^j_aN^k-N(\hbox{det}B)^{1/2}\sqrt{\hbox{det}\Psi}(B^{-1})^b_i(\Psi^{-1}\Psi^{-1})^{ba};\nonumber\\
\dot{\Psi}_{ae}=-A^b_0\bigl(f_{abc}\Psi_{ce}+f_{ebc}\Psi_{ac}\bigr)+N^kD_k\Psi_{ae}\nonumber\\
+N(\hbox{det}B)^{1/2}\sqrt{\hbox{det}\Psi}\bigl[(\hbox{det}B)^{-1}\epsilon^{efd}(\Psi^{-1}\Psi^{-1})^{fb}B^k_dD_k\Psi_{ab}\bigr].
\end{eqnarray}
\noindent
Comparison of (\ref{CANON28}) with (\ref{DERIVATION}) and (\ref{DERIVATION6}) confirms that (\ref{CANON}) is indeed an action for GR for $(\hbox{det}B)$ and $(\hbox{det}\Psi)$ nonzero.  
Note that the second equation of (\ref{CANON28}) is valid if and only if $\Psi_{[ae]}=0$, the consistency of which we can check by examining its antisymmetric part.  Contracting this equation 
with $\epsilon_{gae}$ yields $\epsilon_{gae}\dot{\Psi}_{ae}$ for the left hand side.  The right hand side splits into two terms which we will in turn analyse.  The term $N^k\partial_k(\epsilon_{gae}\Psi_{ae})$ in the covariant derivative is automatically zero when $\Psi_{[ae]}$ is zero.  The antisymmetric part of all the gauge transformation terms is of the form (defining $\eta^b=A^b_0-N^kA^b_k$) 
\begin{eqnarray}
\label{ANTISYM}
\epsilon_{dae}\bigl(f_{abc}\Psi_{ce}+f_{ebc}\Psi_{ac}\bigr)\eta^b\nonumber\\
=\bigl(\bigl(\delta_{eb}\delta_{dc}-\delta_{ec}\delta_{db}\bigr)\Psi_{ce}+\bigl(\delta_{bd}\delta_{ca}-\delta_{ba}\delta_{cd}\bigr)\Psi_{ac}\bigr)\eta^b=2\Psi_{[bd]}\eta^b
\end{eqnarray}
\noindent
which is also antisymmetric.  This leaves remaining the term involving $N$, whose antisymmetric part up to multiplicative factors is
\begin{eqnarray}
\label{ANTISYM1}
\epsilon_{gae}\epsilon^{efd}(\Psi^{-1}\Psi^{-1})^{fc}B^k_dD_k\Psi_{ab}
=\bigl(\delta^f_g\delta^d_a-\delta^f_a\delta^d_g\bigr)(\Psi^{-1}\Psi^{-1})^{fb}B^k_dD_k\Psi_{ab}\nonumber\\
=(\Psi^{-1}\Psi^{-1})^{gb}B^k_aD_k\Psi_{ab}-(\Psi^{-1}\Psi^{-1})^{fb}B^k_gD_k\Psi_{fb}.
\end{eqnarray}
\noindent
Note that the first term on the right hand side of (\ref{ANTISYM1}) is directly proportional to the Gauss' constraint.  The second term can be written as
\begin{eqnarray}
\label{ANTISYM2}
(\Psi^{-1}\Psi^{-1})^{fb}B^k_gD_k\Psi_{fb}
=B^k_d\partial_k(\Lambda+\hbox{tr}\Psi^{-1})
\end{eqnarray}
\noindent  
when $\Psi_{ae}$ is symmetric.  Note that we have appended $\Lambda$ as a constant of spatial integration.  This permits the identification of the covariant derivative with the spatial derivative of the Hamiltonian constraint which is a gauge scalar.  This term also vanishes on-shell.  The result is that the previous manipulations involving $\Psi_{[ae]}$ are valid when all initial value constraints hold, namely on the constraint surface.\par 
\indent 
We have obtained time evolution equations (\ref{CANON28}) without the use of Poisson brackets.  Using these results, we can deduce what the appropriate brackets should be.  It will suffice to deduce them using the evolution induced solely by the Hamiltonian constraint.  These are given weakly by
\begin{eqnarray}
\label{CANON29}
\delta_NA^a_i=\{A^a_i,H[N]\}^{*}=-N(\hbox{det}B)^{1/2}\sqrt{\hbox{det}\Psi}(\Psi^{-1}\Psi^{-1})^{eb}\{A^a_i,\Psi_{be}\}^{*};\nonumber\\
\delta_N\Psi_{ae}=\{\Psi_{ae},H[N]\}^{*}=-N(\hbox{det}B)^{1/2}\sqrt{\hbox{det}\Psi}(\Psi^{-1}\Psi^{-1})^{fb}\{\Psi_{ae},\Psi_{bf}\}^{*}.
\end{eqnarray}
\noindent
Using the fact that the brackets between components of the connection are zero (since there were no ${\delta\Psi}\wedge{\delta\Psi}$ terms in the previous attempt to define a symplectic matrix) and by comparison of (\ref{CANON29}) with the $N$ terms of (\ref{CANON28}), we can read off the following fundamental brackets, assuming that the chain rule holds, 
\begin{eqnarray}
\label{CANON30}
\{A^a_i(x),A^b_j(y)\}^{*}=0;~~\{A^a_i(x),\Psi_{be}(y)\}^{*}=\delta^a_b(B^{-1}(y))^e_i\delta^{(3)}(x,y);\nonumber\\
\{\Psi_{ae}(x),\Psi_{bf}(y)\}^{*}=-\epsilon_{efg}(B^i_gD_i\Psi_{ab})(\hbox{det}B)^{-1}.
\end{eqnarray}
\noindent
There are a few observations that can be made regarding (\ref{CANON30}).  First, note that the relations between $A$ and $\Psi$ are noncanonical due to the factor of $B^{-1}$, which inherits its coordinate dependence from $\Psi$.  Secondly, while $\{A,A\}^{*}=0$, it happens that $\{\Psi,\Psi\}^{*}$ is in general nonzero.  Thirdly, the relations might be modifiable by an aribtrary term whose contraction with $\Psi^{-1}\Psi^{-1}$ is zero.  So we do not necessarily claim that (\ref{CANON30}) are the unique relations.  However they do appear to be the simplest relations consistent with the Hamiltonian constraint part of the Lagrange's equations, and in this paper we will use them as a starting result.\par 
\indent
The remainder of this paper will be devoted to performing various consistency checks on (\ref{CANON30}).  We have already obtained (\ref{CANON30}) via transformations of the basic variables under evolution generated by the Hamiltonian constraint $H$, and one can similarly read off from (\ref{CANON28}) the transformations under the Gauss' law and vector constraints $\vec{G}$ and $\vec{V}$.\footnote{A more robust check would be to show that there is no contradiction between these transformations and (\ref{CANON30}).  It is shown in Appendix A that this is indeed the case, namely that (\ref{CANON30}), and (\ref{CANON30}) alone, are sufficient to fix the correct transformation properties of $\Psi_{ae},A^a_i$ under $\vec{G}$ and $\vec{V}$.  This demonstration additionally provides the opportunity to verify that the brackets satisfy the Leibniz rule, which is used repeatedly.}

\section{Consistency checks of the brackets}

We have deduced a bilinear bracket operation which produces the correct Lagrange equations for $I_{Inst}$ and induces the correct behavior of the basic fields under transformations generated by the initial value constraints.  Our remaining task is to assess the Poisson structure, if any, for $I_{Inst}$.  The basic relations are repeated here for continuity
\begin{eqnarray}
\label{ONEFORM10}
\{A^a_i(x),A^b_j(y)\}^{*}=0;~~
\{A^a_i(x),\Psi_{bg}(y)\}^{*}=\delta^a_b(B^{-1})^g_i\delta^{(3)}(x,y);\nonumber\\
\{\Psi_{bf}(x),\Psi_{cg}(y)\}^{*}=-(\hbox{det}B)^{-1}\epsilon^{fgd}(B^j_dD_j\Psi_{bc})\delta^{(3)}(x,y).
\end{eqnarray}
\noindent
We will check (\ref{ONEFORM10}) against the properties of the Poisson bracket and verify which requirements are satisfied.\par 
\indent
The condition of bilinearity can be taken to be self-evident from the manipulations of this paper thus far.  The condition of antisymmetry is trivially satisfied for the first two relations of (\ref{ONEFORM10}), which leaves the third relation.  Let us perform the relabelling $b\leftrightarrow{c}$ and $f\leftrightarrow{g}$ on these latter terms
\begin{eqnarray}
\label{ONEFORM11}
\{\Psi_{cg}(x),\Psi_{bf}(y)\}^{*}=-(\hbox{det}B)^{-1}\epsilon^{gfd}(B^j_dD_j\Psi_{cb})\delta^{(3)}(x,y).
\end{eqnarray}
\noindent
The sum of (\ref{ONEFORM11}) and the third relation of (\ref{ONEFORM10}) is $2D_j\Psi_{[bc]}$, whereas for Poisson brackets this sum must vanish.  This implies that we must have $D_j\Psi_{[bc]}=0$, which holds on the constraint surface.  (ii) The Leibniz rule has already been verified in checking the transformation properties of $A,\Psi$ in Appendix A.  (iii) This leaves remaining the Jacobi identity.  Checking it directly would be tedious, and so we will rather proceed by relation to the analogously associated symplectic structure as follows.\par 
\indent 
First we will use the relations (\ref{ONEFORM10}) to construct a matrix $P^{IJ}$, given by
\begin{displaymath}
P^{IJ}=
\left(\begin{array}{cc}
\{A^a_i,A^c_k\}^{*} & \{A^a_i,\Psi_{bf}\}^{*}\\
\{\Psi_{ae},A^c_k\}^{*} & \{\Psi_{ae},\Psi_{bf}\}^{*}\\
\end{array}\right)
\end{displaymath}
\begin{displaymath}
=
\left(\begin{array}{cc}
0 & \delta^a_b(B^{-1})^f_i\\
-\delta^c_a(B^{-1})^e_k & -(\hbox{det}B)^{-1}\epsilon_{efg}B^m_gD_m\Psi_{ab}\\
\end{array}\right)
\delta^{(3)}(x,y).
\end{displaymath}
Then the strategy is to find $(P^{-1})_{IJ}$ and see to which extent this defines a symplectic matrix $\boldsymbol{\Omega}$.  This can be found via a relation of the form
\begin{displaymath}
\left(\begin{array}{cc}
0 & \beta\\
-\beta & \alpha\\
\end{array}\right)
^{-1}=
\left(\begin{array}{cc}
 \beta^{-1}\alpha\beta^{-1} & -\beta^{-1}\\
\beta^{-1} & 0\\
\end{array}\right),
\end{displaymath}
\noindent
where $\beta$ and $\alpha$ are block nine by nine matrices, with $\hbox{det}\beta\neq{0}$.  Using this relation, then we have
\begin{displaymath}
(P^{-1})_{IJ}=
\left(\begin{array}{cc}
\epsilon^{imk}D_m\Psi_{ab} & -\delta^a_bB^i_f\\
\delta^a_cB^k_e & 0\\
\end{array}\right)
\delta^{(3)}(x,y),
\end{displaymath}
\noindent  
where the upper left block matrix comes from the relation
\begin{eqnarray}
\label{COMESFROM}
\beta^{-1}\alpha\beta^{-1}=-(\hbox{det}B)B^k_e(\epsilon_{efg}B^m_gD_m\Psi_{ab})B^i_f=\epsilon^{imk}D_m\Psi_{ab}.
\end{eqnarray}
\noindent
We can then construct a 2-form $P^{-1}={1 \over 2}(P^{-1})_{IJ}{\delta\xi^I}\wedge{\delta\xi^J}$ by contraction with two-forms
\begin{eqnarray}
\label{CONTRACTION}
P^{-1}
={1 \over 2}\int_{\Sigma}d^3x\biggl[\epsilon^{ijk}(D_j\Psi_{ab}){\delta{A}^a_i}\wedge{\delta{A}^b_k}
-\delta^b_aB^i_f{\delta{A}^a_i}\wedge{\delta\Psi_{bf}}
+\delta^a_cB^k_e{\delta\Psi_{ae}}\wedge{\delta{A}^c_k}\biggr].
\end{eqnarray}
\noindent
If $P^{-1}$ defines a symplectic 2-form, then we must have $\delta{P}^{-1}=0$.  Taking the exterior derivative of (\ref{CONTRACTION}), we have
\begin{eqnarray}
\label{CONTRACTION1}
\delta{P}^{-1}={1 \over 2}\int_{\Sigma}d^3x\biggl[\epsilon^{ijk}(D_j\Psi_{ae})\wedge{\delta{A}^a_i}\wedge{\delta{A}^e_k}\nonumber\\
+\epsilon^{imk}\bigl(f_{adc}\Psi_{ce}+f_{edc}\Psi_{ac}\bigr){\delta{A}^d_m}\wedge{\delta{A}^a_i}\wedge{\delta{A}^e_k}\nonumber\\
+2\epsilon^{ijk}D_j(\delta{A}^e_k)\wedge{\delta\Psi_{ae}}\wedge{\delta{A}^a_i}\biggr]
\end{eqnarray}
\noindent
where we have used the Leibniz rule acting on the covariant derivative.  The second term of (\ref{CONTRACTION1}) is a sum of terms of the form
\begin{eqnarray}
\label{CONTRACTION2}
{\delta{A}^{[d}_{[m}}}\wedge{\delta{A}^{a]}_i}\wedge{\delta{A}^e_{k]}
+{\delta{A}^{[e}_{[k}}\wedge{\delta{A}^{d]}_m}\wedge{\delta{A}^a_{i]}}=0
\end{eqnarray}
\noindent
which vanishes due to antisymmetry.  Splitting the last term of (\ref{CONTRACTION1}) into its clones with $k\leftrightarrow{i}$ and $a\leftrightarrow{e}$ on the second term, then (\ref{CONTRACTION1}) reduces to
\begin{eqnarray} 
\label{CONTRACTION3}
\int_{\Sigma}d^3x\biggl[-{\delta\Psi_{ae}}\wedge\epsilon^{ijk}D_j({\delta{A}^a_i}\wedge{\delta{A}^e_k})
+\epsilon^{ijk}D_j{(\delta{A}^e_k)}\wedge{\delta\Psi_{ae}}\wedge{\delta{A}^a_i}\nonumber\\
+\epsilon^{kji}D_j{(\delta{A}^a_i)}\wedge{\delta\Psi_{ea}}\wedge{\delta{A^e_k}}\biggr].
\end{eqnarray}
\noindent
Separating $\Psi_{ae}=\Psi_{(ae)}+\Psi_{[ea]}$ into its symmetric and antisymmetric components we see that the contribution to (\ref{CONTRACTION3}) proportional to $\Psi_{(ae)}$ cancels out.  This is nothing other than $\delta\Omega_{Inst}=0$ with $\Omega_{Inst}$ as in (\ref{ONEFORM7}), which is true since $\Omega_{Inst}$ is the exterior derivative of the one form $\theta_{Inst}$ restricted to symmetric $\Psi_{ae}$.\footnote{So the condition $\delta^2\Omega_{Inst}=0$ holds by construction.}  So we are left with the contribution due to $\Psi_{[ae]}$, given by
\begin{eqnarray}
\label{CONTRACTION4}
\delta{P}^{-1}={1 \over 2}\int_{\Sigma}d^3x\epsilon^{ijk}\biggl[{(D_j\delta\Psi_{[ae]})}\wedge{\delta{A}^a_i}\wedge{\delta{A}^e_k}
+2{D_j(\delta{A}^e_k)}\wedge{\delta\Psi_{[ae]}}\wedge{\delta{A}^a_i}\biggr]=r.
\end{eqnarray}
\noindent
The remainder $r$ depends on the antisymmetric part of $\Psi_{ae}$, which is not zero except on the $\vec{V}$ constraint surface.  Since $\delta{P}^{-1}$ is not zero off the constraint surface, then this means that the Jacobi identity (iii) does not hold for (\ref{ONEFORM10}) off the constraint surface, and we do not have a Poisson structure, but rather an almost Poisson structure.  When $\Psi_{[ae]}=0$, then the almost Poisson structure reduces to a Poisson structure which is not symplectic.  This is because the dimension of the resulting phase space becomes $6+9=15$ which is not even.\par

\section{Conclusion and future research}

We have acquired some insight into the geometric structure of the instanton representation of Plebanski gravity $I_{Inst}$, an action for gravity based on the noncanonical variables $\Psi_{ae}$ and $A^a_i$.  While $I_{Inst}$ does not exhibit a symplectic structure on its full phase space, we have shown the following.  (i) The fundamental brackets (\ref{CANON30}) produce the correct Lagrange's equations modulo terms proportional to the initial value constraints and their spatial derivatives, which vanish on the constraint surface.  This confirms that (\ref{CANON}) is indeed a stand-alone action for GR based upon the variables $\Psi_{ae},A^a_i$. (ii) We have checked that these brackets correctly reproduce the required transformation properties of the variables $A^a_i$ and $\Psi_{be}$ under transformations generated by the initial value constraints.  (iii) The geometric structure of $I_{Inst}$ can be categorized as an almost-Poisson structure which becomes a Poisson structure on the constraint surface of the vector constraint.  Since the main results of this paper have been obtained modulo the initial value constraints and their spatial derivatives, it will be necessary to check that the constraints in the variables of $I_{Inst}$ have a closed Poisson (or perhaps more appropriately, almost-Poisson) algebra.  It has been shown in \cite{EYO} that the initial value constrants are preserved under evolution by the Lagrange equations of motion.  This task we relegate to the next paper \cite{POISSON}, since it was the aim of the present paper merely to determine what the fundamental brackets are.  

\section{Appendix A: Transformation properties of the basic fields}

\subsection{SO(3,C) gauge-transformations}

We will first verify the transformation properties of the basic fields induced by the noncanonical brackets ((\ref{CANON30}), under transformations generated by the smeared Gauss' law constraint
\begin{eqnarray}
\label{THESMEARED}
\vec{G}[\vec{\theta}]=-\int_{\Sigma}d^3x\theta^bB^j_fD_i\Psi_{bf}.
\end{eqnarray} 
\noindent
For transformations generated by (\ref{THESMEARED}) we have
\begin{eqnarray}
\label{CANON31}
\delta_{\vec{\theta}}A^a_i(x)=-\{A^a_i(x),\int_{\Sigma}d^3y\theta^b(y)B^j_f(y)D^y_j\Psi_{bf}(y)\}^{*}\nonumber\\
=\int_{\Sigma}d^3yB^j_f(y)(D^y_j\theta^b(y))\{A^a_i(x),\Psi_{bf}(y)\}^{*}\nonumber\\
=\int_{\Sigma}d^3yB^j_f(y)(D^y_j\theta^b(y))\delta^a_b(B^{-1}(y))^f_i\delta^{(3)}(x,y)=D_i\theta^a(x)
\end{eqnarray}
\noindent
where we have used (\ref{CANON30}) in the last line.  The connection transforms as a gauge potential, which leads to the conclusion that $\vec{G}[\vec{\theta}]$ is the generator of gauge transformations parametrized by $\theta^a$.\par 
\indent
To reaffirm this, we will verify the transformation properties of $\Psi_{ae}$ induced by (\ref{CANON30}).  This is given by
\begin{eqnarray}
\label{CANON32}
-\delta_{\vec{\theta}}\Psi_{bf}(x)=\{\Psi_{bf}(x),\int_{\Sigma}d^3y\theta^a(y)B^i_e(y)D^y_i\Psi_{ae}(y)\}^{*}\nonumber\\
=\int_{\Sigma}d^3y\theta^a(y)\{\Psi_{bf}(x),B^i_e(y)\}^{*}D^y_i\Psi_{ae}(y)\nonumber\\
+\int_{\Sigma}d^3y\theta^a(y)B^i_e(y)\{\Psi_{bf}(x),A^d_i(y)\}^{*}\bigl(f_{adc}\Psi_{ce}(y)+f_{edc}\Psi_{ac}(y)\bigr)\nonumber\\
+\int_{\Sigma}d^3y\theta^b(y)B^i_e(y)D^y_i(\{\Psi_{bf}(x),\Psi_{ae}(y)\}^{*})
\end{eqnarray}
\noindent
where we have used the Leibniz rule.  Equation (\ref{CANON32}) consists three terms which we will evaluate in turn, using (\ref{CANON30}) in conjunction with the appropriate integration by parts.  The first term on the right hand side of (\ref{CANON32}) can be written as
\begin{eqnarray}
\label{CANON33}
\int_{\Sigma}d^3y\theta^a(y)\epsilon^{imn}D_m^y\{\Psi_{bf}(x),A^e_n(y)\}^{*}D^y\Psi_{ae}(y)\nonumber\\
=-\int_{\Sigma}d^3y\theta^a(y)\epsilon^{imn}D^y_m(\delta^e_b(B^{-1}(x))^f_n\delta^{(3)}(x,y))D^y_i\Psi_{ae}(y).
\end{eqnarray}
\noindent
We must integrate (\ref{CANON33}) by parts with respect to $y$, which gets rid of the minus sign arising from the relations (\ref{CANON30}).  Note, as a general rule, that the $B^{-1}$ from (\ref{CANON}) should be immune to differentiation, since it inherits its $x$ dependence from $\Psi_{bf}(x)$.  Also, the covariant derivative $D^y$ can act only on quantities which depend only on $y$.  So upon integrating the delta function and relabelling index $m\rightarrow{k}$, equation (\ref{CANON33}) yields
\begin{eqnarray}
\label{CANON34}
(B^{-1}(x))^f_k\int_{\Sigma}d^3y\epsilon^{ijk}D^y_j(\theta^a(y)D^y_i\Psi_{ab}(y))\delta^{(3)}(x,y)
=(B^{-1})^f_kD_j(\epsilon^{ijk}\theta^aD_i\Psi_{ab}).
\end{eqnarray}
\noindent
The second term on the right hand side of (\ref{CANON32}), using (\ref{CANON30}), is
\begin{eqnarray}
\label{CANON35}
-\int_{\Sigma}d^3y\theta^a(y)B^i_e(y)(\delta^d_b(B^{-1}(x))^f_i\delta^{(3)}(x,y))\bigl(f_{adc}\Psi_{ce}(y)+f_{edc}\Psi_{ac}(y)\bigr)\nonumber\\
=-\theta^a\bigl(f_{abc}\Psi_{cf}+f_{fbc}\Psi_{ac}\bigr)
\end{eqnarray}
\noindent
where we have relabelled $e\leftrightarrow{f}$ and $d\leftrightarrow{b}$ on the second term.  From the third term of (\ref{CANON32}) there is a cancellation of one minus sign from the $\{\Psi,\Psi\}$ relations of (\ref{CANON30}) with another minus sign from the integration by parts.  Then applying the Bianchi identity $D_iB^i_a=0$, we have
\begin{eqnarray}
\label{CANON36}
\int_{\Sigma}d^3y(B^i_e(y)D^y_i\theta^a(y))\epsilon_{feg}(B^j_g(x)D^x_j\Psi_{ba}(x)(\hbox{det}B(x))^{-1}\delta^{(3)}(x,y)\nonumber\\
=\epsilon^{ijk}(B^{-1})^f_k(D_i\theta^a)(D_j\Psi_{ba}),
\end{eqnarray}
\noindent
where we have also used the property of determinants of 3 by 3 matrices.  Combining the results of (\ref{CANON34}), (\ref{CANON35}) and (\ref{CANON36}) we have
\begin{eqnarray}
\label{CANON37}
-\delta_{\vec{\theta}}\Psi_{bf}=(B^{-1})^f_kD_j(\epsilon^{ijk}\theta^aD_i\Psi_{ab})
-\theta^a\bigl(f_{abc}\Psi_{cf}+f_{fbc}\Psi_{ac}\bigr)\nonumber\\
+\epsilon^{ijk}(B^{-1})^f_k(D_i\theta^a)(D_j\Psi_{ba}\nonumber\\
=\epsilon^{ijk}(B^{-1})^f_k(D_j\theta^a)(D_i\Psi_{ab})+(B^{-1})^f_k\theta^a\epsilon^{ijk}D_jD_i\Psi_{ab}\nonumber\\
-\theta^a\bigl(f_{abc}\Psi_{cf}+f_{fbc}\Psi_{ac}\bigr)+\epsilon^{ijk}(B^{-1})^f_k(D_i\theta^a)(D_j\Psi_{ba})
\end{eqnarray}
\noindent
where we have used the Leibniz rule on the first term.  A useful relation is the definition of curvature as the commutator of covariant derivatives
\begin{eqnarray}
\label{CANON38}
\epsilon^{ijk}D_jD_kv_a=f_{abc}B^i_bv_c;~~\epsilon^{ijk}D_jD_k\Psi_{ab}=B^i_g\bigl(f_{agc}\Psi_{cb}+f_{bgc}\Psi_{ac}\bigr).
\end{eqnarray}
\noindent
Equation (\ref{CANON38}) tacitly assumes that $\Psi_{ae}$ behaves as a $SO(3,C)$ tensor of second rank, which is what we are attempting to prove in the first place.  This assumption will have been justified by the fact that it will not lead to any contradictions in the final result.  Applying (\ref{CANON38}) to the second term on the right hand side of (\ref{CANON37}), we have
\begin{eqnarray}
\label{CANON39}
-\delta_{\vec{\theta}}\Psi_{bf}=\epsilon^{ijk}(B^{-1})^f_k(D_i\theta^a)D_j(\Psi_{ba}-\Psi_{ab})\nonumber\\
-\theta^a\bigl(f_{afc}\Psi_{cb}+f_{bfc}\Psi_{ac}+f_{abc}\Psi_{cf}+f_{fbc}\Psi_{ac}\bigr).
\end{eqnarray}
\noindent
The first term of (\ref{CANON39}) involves the antisymmetric part of $\Psi_{ab}$ and its derivatives, which vanishes on the constraint surface.  Also, the second and fourth terms in brackets multiplying $\theta^a$ cancel, 
yielding $-\delta_{\theta}\Psi_{bf}=2\theta^af_{(bac}\Psi_{cf)}$ which is the correct transformation property of a 
second-rank $SO(3,C)*SO(3,C)$ tensor under $SO(3,C)$ transformations.  Including (\ref{CANON31}), we see that the transformation properties of the basic fields generated by Gauss' law induced by the relations (\ref{CANON30}) are
\begin{eqnarray}
\label{CANON40}
\delta_{\vec{\theta}}A^a_i=D_i\theta^a;~~\delta_{\vec{\theta}}\Psi_{bf}=-\theta^a\bigl(f_{bac}\Psi_{cf}+f_{fac}\Psi_{cb}\bigr).
\end{eqnarray}
\noindent
The connection $A^a_i$ transforms as a gauge potential and $\Psi_{ae}$ transforms as a second-rank $SO(3,C)$ tensor, justifying (\ref{CANON38}).  Hence we have justifiably referred to $\vec{G}$ as the Gauss' law constraint, the generator of $SO(3,C)$ gauge transformations of the basic fields.  In order for the correct $SO(3,C)$ transformation properties to be obtained then the Leibniz rule, which we have used several times in this section, must have been valid for brackets (\ref{CANON30}).  

\subsection{Spatial diffeomorphisms}

We will now establish the transformation properties of the basic variables generated by the smeared vector constraint
\begin{eqnarray}
\label{SMEAREDVECTOR}
\vec{V}[\vec{N}]=\int_{\Sigma}d^3y\epsilon_{ijk}N^iB^j_eB^k_a\Psi_{ae},
\end{eqnarray}
\noindent
as induced by the relations (\ref{CANON30}).  For the connection we have
\begin{eqnarray}
\label{CANON41}
\delta_{\vec{N}}A^d_m(x)=\{A^d_m(x),\int_{\Sigma}d^3y\epsilon_{ijk}N^i(y)B^j_e(y)B^k_a(y)\Psi_{ae}(y)\}^{*}\nonumber\\
=\int_{\Sigma}d^3yN^i(y)B^j_e(y)B^k_a(y)\{A^d_m(x),\Psi_{ae}(y)\}^{*}\nonumber\\
=\delta^d_a(B^{-1}(x))^e_m\int_{\Sigma}d^3y\epsilon_{ijk}N^i(y)B^j_e(y)B^k_a(y)\delta^{(3)}(x,y)
=\epsilon_{mki}B^k_dN^i.
\end{eqnarray}
\noindent
Equation (\ref{CANON41}) is not strictly speaking a spatial diffeomorphism, but rather a spatial diffeomorphism with parameter $N^i$ corrected by a $SO(3,C)$ gauge transformation with field-dependent parameter $\eta^a=N^iA^a_i$.\par 
\indent
Let us now move on to the transformation of $\Psi_{bf}$.  This is given by
\begin{eqnarray}
\label{CANON42}
\delta_{\vec{N}}\Psi_{bf}(x)=\{\Psi_{bf}(x),\int_{\Sigma}d^3y\epsilon_{ijk}
N^i(y)B^j_e(y)B^k_a(y)\Psi_{ae}(y)\}^{*}\nonumber\\
=\epsilon_{ijk}N^i(y)\{\Psi_{bf}(x),B^j_e(y)\}^{*}B^k_a(y)
\Psi_{ae}\nonumber\\
+\epsilon_{ijk}N^i(y)B^j_e(y)\{\Psi_{bf}(x),B^k_a(y)\}^{*}
\Psi_{ae}(y)
+\epsilon_{ijk}N^i(y)B^j_e(y)B^k_a(y)
\{\Psi_{bf}(x),\Psi_{ae}(y)\}^{*}
\end{eqnarray}
\noindent
where we have applied the Leibniz rule.  Proceeding along, we have
\begin{eqnarray}
\label{WEGET}
\int_{\Sigma}d^3xN^i(y)\biggl[\epsilon_{ijk}\epsilon^{jmn}D_m^y\{\Psi_{bf}(x),A^e_n(y)\}^{*}\Psi_{ae}(y)\nonumber\\
+B^j_e(y)\epsilon_{kmn}D_m^y\{\Psi_{bf}(x),A^a_n(y)\}^{*}\Psi_{ae}(y)
+\epsilon_{ijk}B^j_e(y)B^k_a(y)\{\Psi_{bf}(x),\Psi_{ae}(y)\}^{*}\biggr].
\end{eqnarray}
\noindent 
Then applying the relations (\ref{CANON30}), integrating by parts and the use of epsilon identities,\footnote{Recal that the $B^{-1}$ factor from the $\{A,\Psi\}$ relation inherits its x dependence from $\Psi$, and thereform should remain undifferentiated by $D^y_m$.} (\ref{WEGET}) becomes
\begin{eqnarray}
\label{CANON44}
(B^{-1})^f_n\bigl(\delta^m_k\delta^n_i-\delta^m_i\delta^n_k\bigr)D_m(N^iB^k_a\Psi_{ab})\nonumber\\
+(B^{-1})^f_n\bigl(\delta^m_i\delta^n_j-\delta^m_j\delta^n_i\bigr)D_m(N^iB^j_e\Psi_{be})
-N^i(B^{-1})^d_i\epsilon_{dea}\epsilon_{feg}B^m_gD_m\Psi_{ba}\nonumber\\
=(B^{-1})^f_nD_m\bigl(N^nB^m_e\Psi_{eb}-N^mB^n_e\Psi_{eb}
+N^mB^n_e\Psi_{be}-N^nB^m_e\Psi_{be}\bigr)\nonumber\\
-N^i(B^{-1})^d_i\bigl(\delta_{df}\delta_{ag}-\delta_{dg}\delta_{af}\bigr)B^m_gD_m\Psi_{ba}\nonumber\\
=4(B^{-1})^f_nD_m(N^{[n}B^{m]}_e\Psi_{[eb]})-N^i(B^{-1})^f_iB^m_aD_m\Psi_{ba}+N^iD_i\Psi_{bf}.
\end{eqnarray}
\noindent
The first and second terms on the right hand side of (\ref{CANON44}) are proportional to the $\vec{V}$ and $\vec{G}$ constraints.  So on the constraint surface, the transformation reduces to the last term which is a spatial diffeomorphism parametrized by $N^i$, corrected by a gauge transformation with field-dependent parameter $N^iA^a_i$.  So we have the following results
\begin{eqnarray}
\label{CANON45}
\delta_{\xi}A^a_i=\epsilon_{ijk}B^j_aN^k;~~\delta_{\vec{N}}\Psi_{bf}=N^kD_k\Psi_{bf}.
\end{eqnarray}
\noindent 
The basic variables transform as expected under the vector constraint, namely via spatial diffeomorphism corrected by a $SO(3,C)$ gauge transformation with field-dependent parameter $\eta^a=N^kA^a_k$.  This confirms the predictions from the Lagrange equations.

\end{document}